\documentclass[twocolumn]{aa}  

\usepackage{graphicx}
\usepackage{txfonts}
\usepackage{hyperref}
\usepackage{natbib}
\usepackage{float}
\usepackage{enumitem}
\usepackage{amsmath}
\usepackage{amssymb}
\usepackage{layout}
\usepackage{booktabs}
\usepackage{textgreek}
\usepackage[switch]{lineno}
\usepackage{orcidlink}
\usepackage{overpic}

\bibpunct{(}{)}{;}{a}{}{,}
\usepackage{xcolor}
\definecolor{darkblue}{cmyk}{55,17,0,0}
\hypersetup{colorlinks=true,
  linkcolor=darkblue,
  linkbordercolor=darkblue,
  citecolor=darkblue,
  urlcolor=darkblue
  }

\definecolor{al}{rgb}{0.6,0.2,0.0}
\definecolor{as}{rgb}{0.4,0.2,0.3}
\definecolor{mvn}{rgb}{0.5,0.5,0.9}

\usepackage[normalem]{ulem} 
\newcommand{\sophi}{{\sc SO/PHI}}
\newcommand{\hrt}{{\sc SO/PHI-HRT}}
\newcommand{\hmi}{{\sc SDO/HMI}}

\newcommand{\degree}{$^\circ$}
\makeatletter
\renewcommand*\aa@pageof{, page \thepage{} of \pageref*{LastPage}}

\begin{document} 

   \title{Solar photospheric velocities measured in space: a comparison between SO/PHI-HRT and SDO/HMI}
   
   \author{D.~Calchetti\inst{1}\fnmsep\thanks{\hbox{Corresponding author: Daniele~Calchetti} \hbox{\email{calchetti@mps.mpg.de}}}\orcidlink{0000-0003-2755-5295}
     \and
   K.~Albert\inst{1}\orcidlink{0000-0002-3776-9548} \and
   F.~J.~Bail{\'e}n\inst{2,3}\orcidlink{0000-0002-7318-3536} \and
   J.~Blanco~Rodr\'\i guez\inst{4,3}\orcidlink{0000-0002-2055-441X} \and
   J.~S.~Castellanos Dur\'an\inst{1}\orcidlink{0000-0003-4319-2009} \and
   A.~Feller\inst{1}\orcidlink{0009-0009-4425-599X} \and
   A.~Gandorfer\inst{1}\orcidlink{0000-0002-9972-9840} \and
   J.~Hirzberger\inst{1} \and
   J.~Sinjan\inst{1}\orcidlink{0000-0002-5387-636X} \and
   X.~Li\inst{1}\orcidlink{0000-0001-8164-5633} \and
   T.~Oba\inst{5}\orcidlink{0000-0002-7044-6281} \and
   D.~Orozco~Su\' arez\inst{2,3}\orcidlink{0000-0001-8829-1938} \and
   T.~L.~Riethm{\"u}ller\inst{1}\orcidlink{0000-0001-6317-4380} \and
   J.~Schou\inst{1}\orcidlink{0000-0002-2391-6156} \and
   S.~K.~Solanki\inst{1}\orcidlink{0000-0002-3418-8449}  \and 
   H.~Strecker\inst{2,3}\orcidlink{0000-0003-1483-4535} \and
   A.~Ulyanov\inst{1} \and
   G.~Valori\inst{1}\orcidlink{0000-0001-7809-0067}
   }

   \institute{
         Max-Planck-Institut f\"ur Sonnensystemforschung, Justus-von-Liebig-Weg 3, 37077 G\"ottingen, Germany
         \and
         Instituto de Astrofísica de Andalucía (IAA-CSIC), Apartado de Correos 3004, E-18080 Granada, Spain
         \and
         Spanish Space Solar Physics Consortium (S$^{3}$PC)
         \and
         Universitat de Val\`encia, Catedr\'atico Jos\'e Beltr\'an 2, E-46980 Paterna-Valencia, Spain
         \and
         Advanced Research Center for Space Science and Technology, Institute of Science and Engineering, Kanazawa University, 920-1192 Kakuma-machi, Kanazawa, Ishikawa, Japan
    }

\date{Received ***; accepted ***}
 
  \abstract
  {The Polarimetric and Helioseismic Imager (SO/PHI) onboard Solar Orbiter is a spectropolarimeter scanning the Fe~{\sc i} line at 617.3 nm, providing data of the solar photosphere. The same line is sampled by the Helioseismic and Magnetic Imager (HMI) on board the Solar Dynamics Observatory (SDO) and many other on-ground instruments.
In this paper, we aim at assessing the consistency between line-of-sight (LoS) velocity measurements from the two instruments. Reliable measurements of up and down flows from SO/PHI are crucial and unique when Solar Orbiter is facing the far side of the Sun. Also, a combination of measurements from two vantage points to study horizontal flows must rely on consistent observations.
For this purpose, we compare the LoS velocity measured by SO/PHI's High Resolution Telescope (SO/PHI-HRT) and SDO/HMI on 29 March 2023, when Solar Orbiter was crossing the Sun-Earth line at a distance of 0.39~au from the Sun. 
Because such co-alignments are rare, this configuration offered an almost unique opportunity to directly compare data products from both telescopes. 
The data are aligned and remapped to allow a pixel-by-pixel comparison of the whole time series of 4 hours length. Temporal and spatial variations are considered for a direct combination of the measurements. 
The LoS velocity distributions are evaluated and a clear linear relation is found between the two instruments with a slope of 0.96 and a correlation of 92\%. 
We find that the signals form at similar heights, with a separation of 9$\pm$12~km, which is larger than previous estimates. 
A close-up look at the penumbra of a sunspot and its Evershed flow is presented. 
We conclude that the signals inferred by \hrt\ and \hmi\ show very good agreement and high correlation when instrumental effects and large-scale velocities on the Sun are properly accounted for. }

   \titlerunning{Velocity comparison between \hrt\ and \hmi}
   \authorrunning{Calchetti D. et al.}
   \maketitle

\section{Introduction}\label{sec:intro}
The Polarimetric and Helioseismic Imager \citep[\sophi,][]{2020A&A...642A..11S} onboard Solar Orbiter \citep{solo} is the first spectropolarimetric imager to observe the Sun from a vantage point lying outside the Sun-Earth line. 
Two telescopes are part of the instrument, a Full Disk Telescope (SO/PHI-FDT) and a High Resolution Telescope \citep[\hrt,][]{gandorfer2018high}, which both provide measurements of the photospheric magnetic field vector, continuum intensity and line-of-sight (LoS) velocity in the solar photosphere. 
Only a few other satellites have instruments capable of retrieving these photospheric quantities, such as the Spectropolarimeter instrument \citep[SP,][]{2013SoPh..283..579L} of the Solar Optical Telescope \citep[SOT,][]{2008SoPh..249..167T} onboard Hinode \citep{2007SoPh..243....3K}, the Helioseismic and Magnetic Imager onboard the Solar Dynamics Observatory \citep[\hmi,][]{2012SoPh..275..229S,2012SoPh..275....3P}, and the Full-Disk Vector MagnetoGraph onboard the Advanced Space-Based Solar Observatory \citep[ASO-S/FMG,][]{2024SoPh..299..157B,2023SoPh..298...68G}.

Despite the clear advantages of satellite observations, such as the absence of the disturbance from the Earth's atmosphere and the capability of uninterrupted data acquisition, many challenges have to be faced to achieve a consistent data reduction. 
In the case of \sophi\, the highly eccentric orbit and low telemetry periods when the spacecraft is at more than~1\,au from Earth, make the \sophi\ data reduction considerably more challenging. 
On the other hand, \sophi\ can for the very first time measure the photospheric properties of the Sun from different viewpoints than Earth-bound observatories \citep{solo}. 
This unique orbit opens the possibility for novel scientific applications, such as building fast synoptic maps of the LoS magnetic field \citep{2024A&A...681A..59L,2024A&A...682A.108L} or direct combination of intensity and magnetic field data in stereoscopic configuration \citep[e.g.][]{2023A&A...677A..25V,2023A&A...678A.163A,2024SoPh..299...41R}. 
Comparisons of the magnetic field vector between \sophi\ and \hmi\ have been carried out by \cite{2023-jonas} and \cite{2024A&A...685A..28M} to assess the similarity between the data products of these two instruments. 
Such inter-instrument comparisons are fundamental for combining data from different vantage points, particularly when tracking features on the far side of the Sun as observed from Earth \citep[e.g.][]{2025A&A...697A.217F}. 

The same approach can be applied to the photospheric flows measured by multiple instruments. 
Dopplergrams are a crucial data product to study the emergence of active regions \citep[e.g.][]{1998A&A...333.1053L}, waves \citep{2023LRSP...20....1J}, or the dynamics of the magnetic field in active regions \citep[e.g.][]{2015ApJ...811...16K}. 
Providing high quality data comparable to other spectropolarimeters will be a key element for studying the evolution of active regions for more than half of the solar rotation and for disentangling horizontal from vertical flows by combining data from different vantage points.

This work focuses on the comparison of the LoS velocity field measured by \hrt\ and \hmi. 
In Section \ref{sec:data} we describe the data used for this work. 
In Section \ref{sec:remap} the method used for the remapping of the data is explained in detail. 
The results are shown in Section \ref{sec:result} and the discussion and summary are presented in Section \ref{sec:discussion}.

\section{Data}\label{sec:data}
The observations analyzed in this work were acquired while Solar Orbiter was in inferior conjunction (along the Sun-Earth line) with Earth on 29 March 2023 during the Solar Orbiter Observing Plan R\_SMALL\_HRES\_HCAD\_RS-burst \citep{2020A&A...642A...3Z}. 
This orbital configuration gives us the opportunity to have Earth-bound and Solar Orbiter instruments observing the Sun from an almost identical viewing angle. 
Here we focus on the comparison of \hrt\ and \hmi. 
Both telescopes scan the Fe~{\sc i} absorption line at 617.3~nm with the same angular resolution, although this translates to different spatial resolutions on the photosphere due to the different distances of the telescopes from the Sun, which varies greatly along Solar Orbiter's orbit. 
Information about the observation employed in this paper can be found in Table \ref{tab:data}, such as the time of the acquisition, the NOAA number of the observed sunspot, cadence, pixel scale, and the position of the spacecraft. 
More information about the technical comparison between the two telescopes is provided in \cite{2023-jonas}. 

\begin{figure}
    \centering
    \includegraphics[width=0.985\columnwidth]{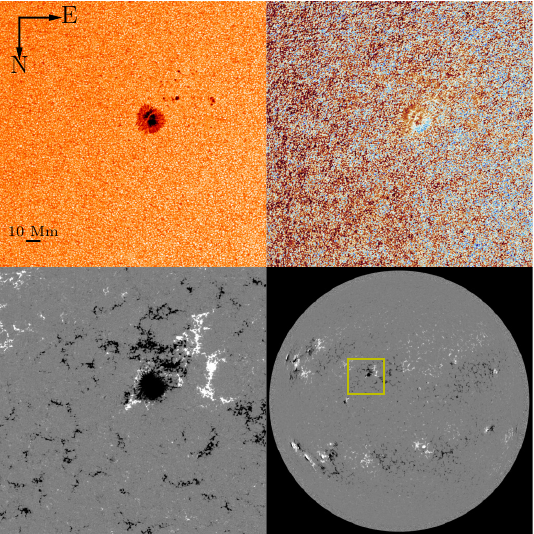}
    \caption{Top left panel: intensity continuum as measured by \hrt. A 10~Mm scale is given as reference. Top right panel: LoS velocity as measured by \hrt. Bottom left panel: LoS magnetic field as measured by \hrt. Bottom right panel: LoS magnetic field as measured by \hmi. The yellow box shows the \hrt\ FoV. The black arrows in the top left panel indicate the solar north (N) and east (E). Note that all the panels of this figure are shown in the \hmi\ detector frame.}
    \label{fig:data}
\end{figure}

An overview of the data is shown in Fig.~\ref{fig:data}. 
The top panels display the continuum intensity and LoS velocity as measured by \hrt, the bottom panel depicts the LoS magnetic field as seen from both instruments, with the \hrt\ FoV indicated by the yellow box in the \hmi\ magnetogram. 
All the panels are shown in the respective detector frame, which means that the \hrt\ $y$ axis is roughly aligned with the north-south direction and \hmi\ is rotated by approximately 180\degree  (as shown by the arrows in the bottom panels). 
\begin{table}[b]
    \caption{Details of the observations. }
    \centering
    \begin{tabular}{lcc}
        \toprule
         & \hrt & \hmi\ \\
         \midrule
         Time & \multicolumn{2}{c}{2023-03-29, 11:40-15:40 UT} \\
        \midrule
        NOAA number & \multicolumn{2}{c}{13262}\\
        \midrule
        $\mu$ angle & \multicolumn{2}{c}{0.96} \\
        \midrule
        Cadence & 60~s & 45~s \\
        \midrule
        Angular separation & \multicolumn{2}{c}{2.3\degree}\\
        \midrule
        Sun distance & 0.39~au & 1~au \\
        \midrule
        Pixel scale & 143~km/px & 362~km/px \\
        \bottomrule
    \end{tabular}
    \tablefoot{The $\mu$ value is defined as the cosine of the heliocentric angle of the center of the field of view (FoV). The pixel scale is defined as at disk center.}
    \label{tab:data}
\end{table}
The data products chosen for this comparison are the 45~s cadence data products of \hmi\ \citep[based on the so-called MDI-like algorithm,][]{2016SoPh..291.1887C} and the newly improved standard results of the \hrt\ pipeline \citep{hrt-pipeline,fatima-PD,2024-fran}, which is now also correcting for residual stray light, based on the inversion of the radiative transfer equations (RTE) in Milne-Eddington approximation with the MILOS code \citep{2007A&A...462.1137O}. 
This data product became the standard data product in most recent data releases\footnote{\href{https://www.cosmos.esa.int/documents/3689933/11863501/SO_PHI-HRT_fifth-data-release_L2_first-version.pdf/43823028-c736-81f4-7ab1-3d29cc4d9a13?t=1768575286145}{Fifth \hrt\ data release notes.}} (see Sinjan et al. under review, for more details). 

We post-process the \hrt\ data with the method described in \cite{2023-calchetti}, which mainly consists of aligning the whole time series and temporally interpolating each frame of the Stokes vector to the same observing time. 
For consistency, instead of a linear interpolation we use an apodized normalized $sinc$ function. 
The choice is motivated by using a similar function to that in the \hmi\ pipeline \citep{sinc-function}, even though no significant dependence is found of the results presented in this work on the employed interpolation scheme. 

\section{Remapping and alignment}\label{sec:remap}
An accurate alignment of the whole time series is needed for a pixel-by-pixel comparison. 
The procedure described below is applied to every \hrt\ observation (240 in total), which is then associated to the closest \hmi\ dataset in time. 
Differences in light travel times are also considered when selecting the \hmi\ dataset. 

Firstly, the World Coordinate System (WCS) of \hrt\ is updated to compensate for three so far unaccounted contributions to the nominal spacecraft pointing: the tracking of the image stabilization system, the translation of the field of view (FoV) due to the change in the focus position, and the pointing inaccuracy of the spacecraft itself. 
Other residuals coming from small variations in the instrument bore-sight are also compensated. 
The alignment is achieved via cross-correlation between the continuum intensity maps of \hrt\ and \hmi, with the latter remapped on the \hrt\ detector frame. 
This procedure is repeated iteratively while updating the WCS information to account for the misalignment found with the cross-correlation until the obtained shift is smaller then a hundredth of a pixel. 

Secondly, we compensate for geometrical distortions in the \hrt\ FoV. 
This step starts by cropping the \hmi\ time series to the desired FoV (see the yellow box in Fig.~\ref{fig:data}) and by tracking the sunspot in the center of the FoV at the Carrington rotation rate, similar to the \textit{mtrack} module in the \hmi\ pipeline. 
We then destretch the \hrt\ LoS magnetic field map, remapped onto the cropped \hmi\ detector frame, with respect to the corresponding \hmi\ magnetogram. 
The pixel-by-pixel shifts obtained by this destretching algorithm are then applied in the algorithm that remaps \hrt\ onto the \hmi\ detector frame with a cubic interpolation. 
Such a procedure can be applied only when the angular separation of the two instruments is close to 0\degree, where differences in viewing geometry are minimized, and is necessary to remove any residual geometrical distortion in the FoV. 

\begin{figure}[t]
    \centering
    \includegraphics[width=0.985\columnwidth]{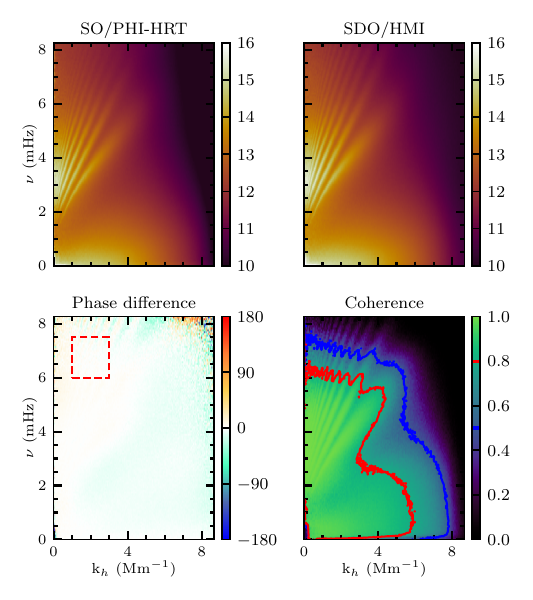}
    \caption{k-$\omega$ diagrams. Top left panel: Power spectrum of the remapped \hrt\ time series (units: $m^2/s^2$). Top right panel: Power spectrum of the \hmi\ time series interpolated to 60~s cadence (units: $m^2/s^2$). Bottom left panel: phase difference between \hrt\ and \hmi\, where positive values means that \hrt\ is lagging behind \hmi\ (units: degrees). The red dashed box shows the area used to compute the height difference between the two signals. Bottom right panel: coherence spectrum between the two signals. The blue and red contours show the 50\% and 80\% coherence respectively.}
    \label{fig:kw}
\end{figure}
Lastly, before remapping the \hrt\ LoS velocity on the \hmi\ image plane, we convolve the \hrt\ Dopplergram with the \hmi\ theoretical point spread function (PSF) and removed the LoS photospheric large scale velocity components (LSC) in the data, including solar differential rotation, spacecraft velocity, meridional flow, convective blueshift, and gravitational redshift \citep[see][for details]{2016-schuck,2021-castellanos}. 
The LSC are also removed from the \hmi\ Dopplergrams. 
More information on the equations used for the LSC are given in Appendix \ref{app:lsc}. 
The final result consists of 240 \hrt\ Dopplergrams of 440$\times$440 pixels remapped onto the \hmi\ detector frame. 
We will refer to this time series from now on as remapped \hrt. 
The cadence of this time series is still 60~s, but the \hmi\ time series has an irregular cadence in order to always consider the closest dataset in time taking into consideration the difference in light travel time. 
Basically, since we consider the 45~s nominal \hmi\ cadence, we excluded one of every four datasets. 
As a consequence, the time difference between the \hrt\ and \hmi\ Dopplergrams alternates between approximately 20, 5, and 10~s repetitively. 

\section{Results}\label{sec:result}
Different types of comparisons are reported in this section. 
Taking advantage of the sub-pixel accuracy in the alignment of the time series, we study the differences between velocities measured by the two instruments in space, time, and in the Fourier space on a pixel by pixel basis. 
Such studies are possible only with an accurate alignment of the data, as described in Sect.~\ref{sec:remap}, which is crucial to reveal both global and localized discrepancies in the data. 

A mask excluding 50 pixels at the edges of the FoV is applied to avoid missing \hrt\ data caused by the alignment and remapping of the time series. 
In addition, we examine the sunspot penumbra region in more detail. 

\subsection{Fourier space}
The comparison between two time series in the Fourier space can be addressed by means of the k-$\omega$ diagrams (see \citealt{2025NRvMP...5...21J} for a description of this method).
Best accuracy is attained by using two time series with the same spatial resolution, cadence, and size. 
The \hmi\ time series constructed for our study has an irregular cadence which has to be corrected for this analysis (see Section \ref{sec:remap}). 
To do that, we reduce the full 45~s cadence \hmi\ observation to the 60~s \hrt\ cadence by removing the highest frequencies in Fourier space, keeping the same starting and ending time of the original series, which means that there is still a (constant) time difference of 20.5~s with respect to the \hrt\ time series. 
This resampled time series is not used for the pixel-by-pixel comparison to avoid any spurious result that might have been generated by this procedure. 

\begin{figure}[t]
    \centering
    \includegraphics[width=0.985\columnwidth]{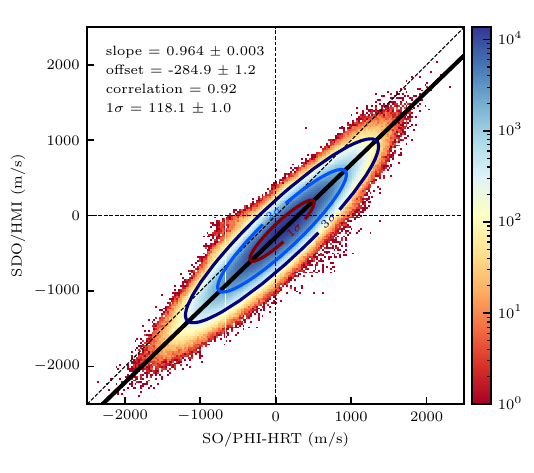}
    \caption{Scatter plot distribution of the velocity field values as measured by the remapped \hrt\ ($x$ axis) and \hmi\ ($y$ axis). The color scale shows the number of points in each bin on a logarithmic scale. The black line shows an orthogonal least square linear fit to the data points, while the contours depict the results of a 2-D Gaussian distribution fit. The slope and offset obtained by the linear fit, the Pearson correlation coefficient, and the 1$\sigma$ spread of the Gaussian fit are indicated in the top left corner of the figure. The uncertainties represent 1~standard deviation estimated using the bootstrap method with 500 sub-samples. }
    \label{fig:scatter-full}
\end{figure}

\begin{figure}[t]
    \centering
    \includegraphics[width=0.985\columnwidth]{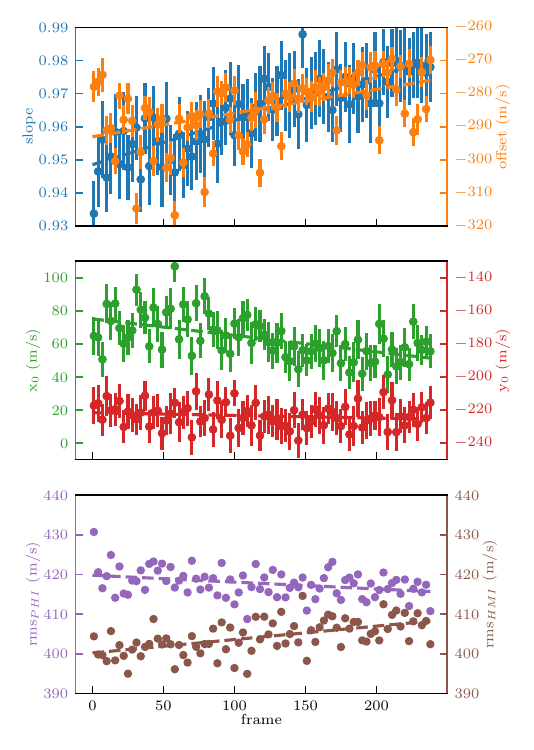}
    \caption{Temporal variation of the fitting results. The dots represent the results of the fits obtained frame by frame, whereas the dashed lines show the linear trends of these results. The error bars represent 1~standard deviation estimated using the bootstrap method with 80 sub-samples used for the fitting. Top panel: slope (left axis, blue) and offset (right axis, orange) obtained by the linear fit. Middle panel: $x$ (left axis, green) and $y$ (right axis, red) coordinates of the mean of the 2-D Gaussian distribution. Bottom panel: root mean square (rms) of the remapped \hrt\ (left axis, purple) and of the \hmi\ (right axis, brown) time series. }
    \label{fig:temporal}
\end{figure}
The power spectra of the remapped \hrt\ and \hmi\ time series and their phase difference and coherence spectrum are shown in Fig.~\ref{fig:kw}. 
The phase difference has been corrected for the time difference between the two time series. 
The coherence spectrum shows the linear relation between two signals and it is not affected by their amplitude \citep{2017ApJ...835..148V}. 
The first thing to notice is the similarity between the power spectra of the two instruments, which is made more evident by the diagrams showing the phase difference and the coherence between the two time series. 
The phase difference remains consistently close to 0$^\circ$ across most of the domain, while the coherence spectrum shows very high values (above 80\%) along the p-modes and f-mode ridges and in the internal gravity waves and granulation regime. 
The phase difference diagrams can be used to estimate the average height difference between the two signals \citep[see equations (2) and (3) in][]{2023-oana}. 
We select the area of the diagram where the waves have real vertical wavenumber, which means that they propagate and they are not evanescent (as when the vertical wave number is imaginary). 
We can then restrict our choice only to vertically propagating waves by considering only small horizontal wavenumbers (see the red dashed box in the bottom right panel in Fig.~\ref{fig:kw}). 
We assume the cut-off frequency, buoyancy frequency, and sound speed to be equal to 5.4~mHz, 4.9~mHz, and 7~km/s respectively. 
We find that the \hrt\ signal forms approximately 9~km above \hmi, with a broad distribution between -10~and~30~km. 
Previous result by \cite{2023-jesper} suggested a formation height of \textasciitilde156~km for \hmi\ and \textasciitilde158~km above the solar surface for \sophi\ obtained with the help of a MURaM simulation. 
This result was obtained by retrieving the velocity through the MDI-like algorithm also for \sophi, so we would expect those signals to be closer than those addressed in that work. 
\begin{figure}[t]
    \centering
    \includegraphics[width=0.84\columnwidth]{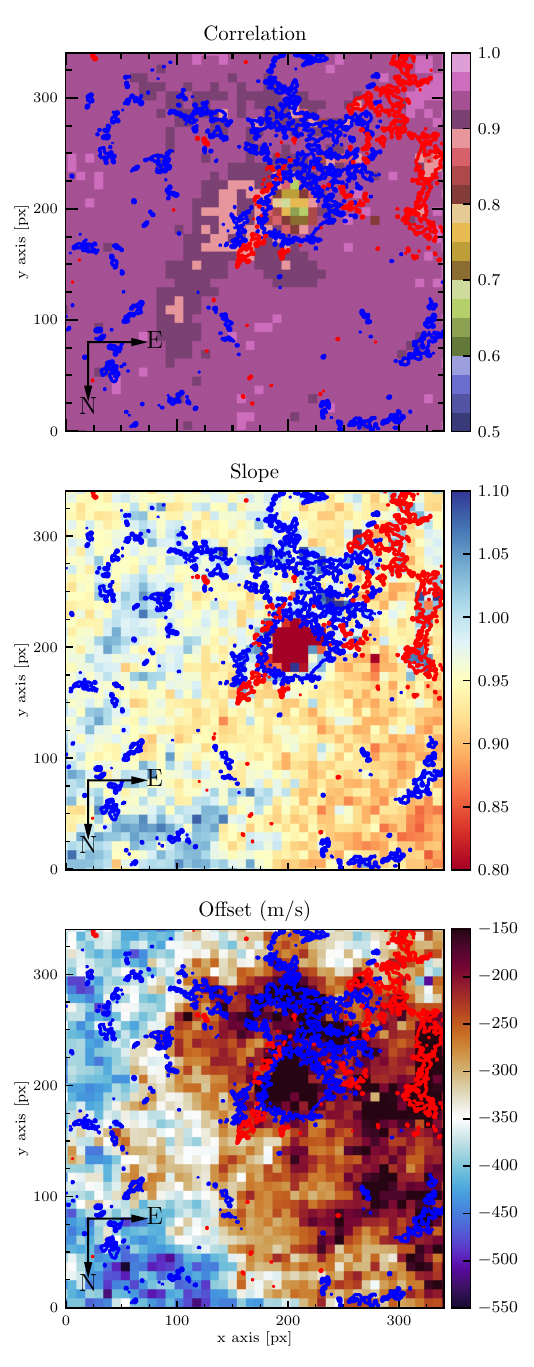}
    \caption{Spatial dependence of the temporal correlation between the remapped \hrt\ and \hmi\ signals (top panel), and of the slope (middle panel) and offset (bottom panel) obtained by the linear fit. The blue and red contours show the -200 and +200~G values of the \hrt\ LoS magnetic field respectively. The FoV is shown as a yellow box in Fig.~\ref{fig:data}. The black arrows in each panel indicate the north (N) and east (E) solar limb as in Fig.~\ref{fig:data}. }
    \label{fig:spatial}
\end{figure}
\subsection{Distributions}
As a second assessment of the similarity of the \hmi\ and \hrt\ Dopplergrams, a scatter plot of the remapped \hrt\ and the non-interpolated \hmi\ time series is shown in Fig.~\ref{fig:scatter-full}. 
Only the datasets with time difference of 5~s are selected for plotting in this figure because the width of the distribution is affected by the time discrepancy. 
The colors correspond to the number of points inside a specific bin in logarithmic scale. 
Two functions have been fit to this distribution: a linear function and a 2-D Gaussian distribution. 
The linear fit shows a clear correspondence between the data, with a slope of 0.964$\pm$0.003, a correlation coefficient of 92\%, and an offset of -284.9$\pm$1.2~m/s. 
The Gaussian distribution indicates the spread between the two signals, which is 118.1$\pm$1.0~m/s at 1$\sigma$. 
The uncertainties are estimated with the bootstrap method by repeating the fitting on 500 random sub-samples. 
The bootstrap is performed by randomly partitioning the data into individual subsets (i.e. sampling without replacement). 
Consequently, every data point appears in only one of the sub-samples, and their union is the full original set of values. 
We also determine the mean of the distribution, which provides information on the offset of the zero point between both time series. 
The average of the distribution for \hmi\ is at -224~m/s, whereas the \hrt\ mean value is at 64~m/s. 
Discussing the absolute calibration of these instruments is a challenge that goes beyond the scope of this work, but it is indeed a point that needs to be mentioned. 
In the \hmi\ case, the reference velocity can be set in many ways available in the literature \citep[e.g.][]{2003A&A...410..695M,2011ARep...55..163G,2009ApJ...697..913O,2011A&A...528A.113D,2013ApJ...765...98W}, including simply setting the mean signal at the center of the Sun to zero (after the removal of all the LSC). 
The same approach is not always possible for \hrt, due to its smaller FoV and the fact that it often points such that solar disk center does not lie in its FoV.  

To investigate the presence of any temporal dependence, we repeated the analysis described above for each of the 240 frames separately. 
Fig.~\ref{fig:temporal} shows a selection of fit parameters as a function of time. 
The uncertainties are estimated with the bootstrap method by repeating the fitting on 80 random sub-samples. 
The top panel displays the results of the linear fit on each of the 240 observation pairs of frames. 
The blue points show the slope and the orange points the offset obtained by the linear fit. 
We can clearly distinguish a linear trend through the 4~hours of observation in which both values evolve towards a value closer to 1. 
The middle panel depicts the coordinates of the average of the Gaussian distribution. 
The left axis (in green, $x$ coordinates of the center of the distribution) corresponds to the mean values of each dataset of the remapped \hrt, the right coordinates (in red, $y$ coordinates of the center of the distribution) to \hmi. 
The range of the two $y$-axes is the same for both coordinates, meaning that the relative variation with time of both values is directly comparable. 
In this case, the \hmi\ values do not show any significant variation with time, while the remapped \hrt\ values go from 70 to 55~m/s. 
The bottom panel shows the root mean square (rms) of the remapped \hrt\ (purple) and \hmi\ (brown) time series, which is used to quantify the amplitudes measured by the instruments. 
The two axes in this case are exactly the same for both distributions. 
The remapped \hrt\ rms is overall higher than \hmi\, as expected by the slope of the linear fit shown in Fig.~\ref{fig:scatter-full} being smaller than one. 
On the other hand, we see a decrease of the rms in \hrt\ of \textasciitilde5~m/s and an increase in \hmi\ of \textasciitilde10~m/s, which is also consistent with the increase of the slope in time. 

In a further step we look at the spatial variation of the results of the linear fit. 
For that, the FoV is divided into square regions of 8$\times$8~pixels and we perform a linear fit in each of these regions through the whole time series. 
We also compute the averaged correlation between the two time series in each sub-region. 
The results are shown in Fig.~\ref{fig:spatial}. 
The top panel shows the spatial-distribution of the correlation, which is overall very high with a mean value of 93\%. 
Lower values are found especially within the sunspot umbra, where the correlation drops to 70-80\%. 
Many factors lead to this low correlation including the low intensity signal, the large line splitting due to the strong magnetic field that can become as large as the total wavelength range in the line that is scanned, the small velocity in the umbra, the few measured wavelength points within the line (which are different for \hmi\ and \sophi), different algorithms to retrieve the velocity signal, all contribute to the low correlations between the two instruments within umbrae (see Sect. \ref{sec:discussion}).  

The middle and bottom panels show the spatial distribution of the slope and offset of the linear fit. 
Here a spatial distribution is quite evident, particularly in the case of the offset. 
The FoV dependence is likely due to calibration residuals, such as an imperfect treatment of the prefilter, etalon cavity correction, or flat correction in \hrt.
The spatial variation does not show any correlation with any of the LSC of either instruments and the amplitude of the signal also does not match with any of the known effects that we compensate for. 
Note that in the middle panel of Fig.~\ref{fig:spatial} low values of the slope are  seen in large parts of the sunspot including parts of the penumbra. 
This is an artifact of the averaging onto larger pixels, which mixes the low (and uncertain) slope values found in the umbra with the high values in the penumbra (see Sect.~\ref{sec:penumbra})

\subsection{Sunspot penumbra}\label{sec:penumbra}
\begin{figure}[t]
  \centering
  \begin{overpic}[width=0.985\columnwidth]{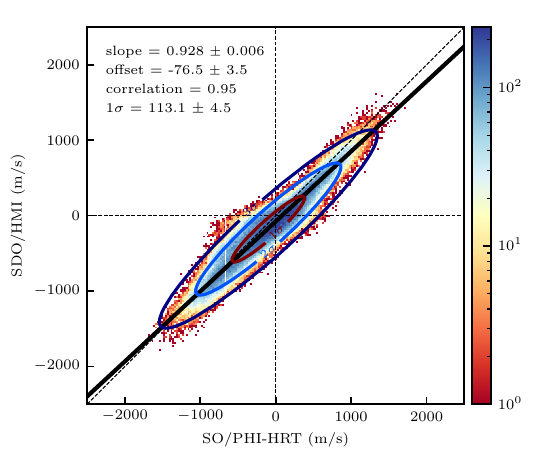}
    \put(57,15){\includegraphics[width=0.14\textwidth]{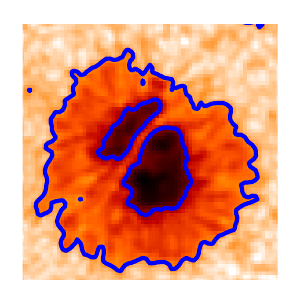}}
  \end{overpic}
  \caption{Scatter plot distribution of the velocity field values as measured by the remapped \hrt\ ($x$ axis) and \hmi\ ($y$ axis) in the sunspot penumbra. The colors and lines are defined as in Fig.~\ref{fig:scatter-full}. The uncertainties represent 1~standard deviation estimated using the bootstrap method with 100 sub-samples. The image in the lower right quadrant shows a snapshot of the sunspot observed by \hmi\ oriented as in Fig.~\ref{fig:data} and the blue contours define the penumbra boundaries. }
  \label{fig:sunspot-scatter}
\end{figure}

In this section the penumbra of the sunspot is analyzed in more detail. 
The penumbra is defined as those pixels with intensity higher than 55\% and lower than 89\% of the averaged intensity continuum in a nearby quiet Sun area. 
A reproduction of the scatter plot between the remapped \hrt\ and the \hmi\ time series, focusing exclusively on the penumbra and on its Evershed flow \citep{1909MNRAS..69..454E} is visible in Fig.~\ref{fig:sunspot-scatter}. 
The contours defining the penumbra boundaries are also visible in the inset. 
The uncertainties are estimated with the bootstrap method by repeating the fitting on 100 random sub-samples. 
It is evident that the scatter of the points is smaller than in Fig.~\ref{fig:scatter-full}, which can be explained by the smaller FoV considered here (and thus smaller spatial variation) and is well represented by the higher correlation coefficient (95\%) with respect to that of the full FoV. 
The slope obtained by the linear fit (0.928$\pm$0.006) is slightly smaller than the one obtained for the whole FoV. 
Additionally, the offset between the two signals is smaller than that found in Fig.~\ref{fig:scatter-full}. 
This is expected from the position of the sunspot in the bottom panel of Fig.~\ref{fig:spatial}, where it lies in an area where the offset is smaller than the average (as obtained by the linear fit shown in Fig.~\ref{fig:scatter-full}). 

\begin{figure*}[t]
    \centering
    \includegraphics[width=0.99\textwidth]{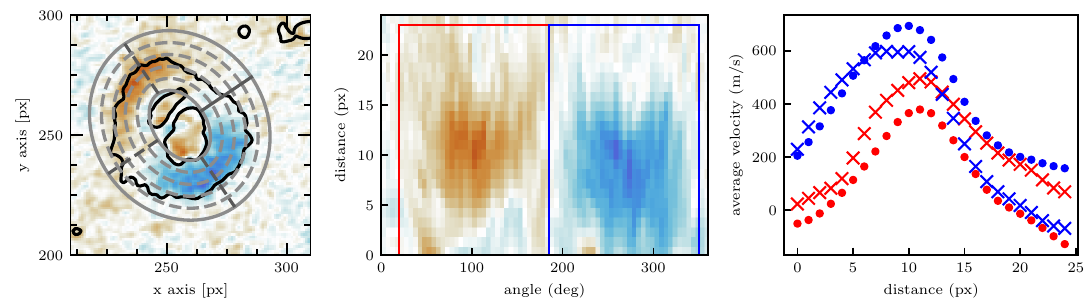}
    \caption{Left panel: remapped \hrt\ LoS velocity signal around the sunspot averaged in time (oriented as in Fig.~\ref{fig:data}). The black contours are at 55\% and 89\% of the average continuum intensity. The gray lines show the grid used for the elliptical polar projection of the penumbra. The solid gray ellipses are the starting and ending point of the radial distance axis, the solid straight gray line shows the starting point of the azimuth angle axis (measured counter-clockwise). Middle panel: same signal as shown in the left panel, but represented on the elliptical polar grid displayed in the left panel. The $x$ axis shows the azimuth, the $y$ axis shows the radial distance (in pixel) from the innermost ellipse. Right panel: azimuthal average of the signal within the blue and red boxes of the middle panel. The crosses show the remapped \hrt\ signal, whereas the points are the \hmi\ signal. The colors correspond to the boxes shown in the middle panel. The values displayed in blue are multiplied by -1 for an easier comparison.}
    \label{fig:sunspot}
\end{figure*}

After assessing that the velocities measured in the penumbra are comparable to those in the rest of the FoV, the comparison of the spatial dependence of the Evershed flow recorded by the two instruments can be addressed. 
Therefore, an elliptical polar grid is defined to encompass the whole penumbra and its Evershed flow, which extends slightly beyond the outer edge of the penumbra. 
The grid is shown as gray lines in the left panel of Fig.~\ref{fig:sunspot}. 
The solid gray ellipses show the starting and ending point of the radial distance. 
The azimuth angle axis is defined starting from the straight solid line in counter-clockwise direction with a $5^\circ$ step size. 

The middle panel of Fig.~\ref{fig:sunspot} displays the velocity on the elliptical polar grid. 
Note that the velocities plotted in this panel are temporal averages over the full 240 frames recorded by the \hrt. 
The two branches of the Evershed flow are clearly visible in the left and right part of the plot, which consists of the limb (outlined by the red box) and disk center (blue box) oriented sides of the penumbra, respectively. 
In order to compare the values obtained by the two instruments, we perform an azimuthal average (along the $x$ axis) in the red and blue boxes for each time series. 
The results are shown in the right panel of Fig.~\ref{fig:sunspot}, where the $x$ axis represents the radial distance on the elliptical polar grid. 
The colors are the same as of the boxes shown in the middle panel, the crosses show the result obtained with the remapped \hrt\ and the circles those from \hmi. 
The curves resulting from the data of the two instruments lie very close to each other until the penumbra-quiet Sun boundary (which lies roughly at a distance of 15~px along the $x$ axis). 
An offset, comparable to that shown in Fig.~\ref{fig:sunspot-scatter}, is visible around the peaks of the velocity signals. 
The asymmetry between the two sides of the penumbra \citep[limb- and disk center-side, e.g.][]{2004A&A...415..731S} is to be expected. 
Also, a difference in the flows can be caused by the asymmetric shape of the sunspot. 

\section{Discussion}\label{sec:discussion}
We have reported a comparison of the LoS velocity measured by \hrt\ and \hmi\ while Solar Orbiter was in inferior conjunction. 
The data acquired by \hrt\ have been post-processed and remapped onto the \hmi\ detector frame to allow a pixel-by-pixel comparison. 
Despite differences in instrumentation, calibration, cadence, and velocity retrieval methods, the comparison shows an excellent agreement of the LoS velocity signals obtained with these two telescopes. 
We measure a height difference of 9$\pm$12~km between the two time series and obtain a slope of 0.94 for the linear fit of the scatter plot between the two instruments and a correlation coefficient of above 90\%. 
The measured height difference is larger than the result obtained from a MURaM simulation by \cite{2023-jesper} and it is possible to estimate the velocity variation due to such a separation.  
If we consider the conservation of mass from one layer to the other, it holds
\begin{equation}
    \rho_{HMI} v_{HMI} = \rho_{PHI} v_{PHI}\,,
\end{equation}
where $\rho$ is the density at each layer and $v$ is the vertical velocity. 
A ratio $v_{HMI}/v_{PHI}$ of 0.952 for 9~km height separation is obtained by considering the atmospheric densities from the VAL-C model \citep{1981ApJS...45..635V}, whereas 6.5~km of separation would be necessary to obtain a ratio equals to 0.964. 
Considering the broad distribution of the height separation between the two instruments, it is not possible to address it as the unique cause for having a ratio smaller than 1. 
One more reason is to be found in the bottom panel of Fig.~\ref{fig:temporal}, which shows that the rms of the remapped \hrt\ is always higher than that of \hmi. 
This is due to the spatial deconvolution of the \hrt\ PSF including the stray light component during the data reduction, a correction that is not included in the \hmi\ pipeline, but it can be made available on request. 
We compare these data with the velocities from \hmi\ without stray light removed, as we want to compare the standard data products, which are the most likely to be used for studies comparing or combing velocities from the two instruments. 
One more element that could also contribute is the different sampling of the spectral line. 
\hrt\ samples the line in five points with a spectral resolution of 58000, \hmi\ in six points with a spectral resolution of 81000. 
\cite{stief-2019} showed that the velocity retrieved with different sampling and at different points along the line can substantially affect the measurement. 
However, a lower spectral resolution reduces the variation in the retrieved velocity near the core of the line (see the yellow curve in Fig.~5 in \citealt{stief-2019} for the \hmi\ case), meaning that we do no expect this to be the main component of the difference between the two instruments. 

The temporal variation of the linear and Gaussian fitting results (presented in Fig.~\ref{fig:temporal}) is mostly due to two factors: a variation of \textasciitilde10~m/s in the average distribution of the \hrt\ velocity field (middle panel) and the opposite variation of the rms of the two instruments (bottom panel). 
The variation of the average value in \hrt\ can be caused by the intrinsic sensitivity of the instrument, due to the voltage applied to the etalon, which translates into \textasciitilde17~m/s, and due to the thermal stabilization, which converts into \textasciitilde19~m/s. 
While the rms increase in \hmi\ is consistent with the findings of \cite{2020ApJ...890..141O}, the slow decrease in \hrt\ is likely due to the optical quality of the data through the 4-hour observation. 
In particular, a major component can be attributed to the focus of the instrument, which changes with time because of the fast approach of the spacecraft to the Sun and the associated thermal drifts. 
A possible indication is the presence of the same decrease in the unreconstructed \hrt\ LoS velocity data and in the intensity contrast in the raw data. 
These different trends in the two instruments' rms are responsible for the variation over time of the slope obtained with the linear fit. 
The increase of the slope up to values close to 1 (and not the opposite behavior) is mostly dependent on the \hrt\ decrease in rms. 
It is also important to point out that the values of the slope are comparable to those of the ratio between the \hmi\ and \hrt\ rms. 
This can indicate a possible strategy for directly combining the velocities measured by these two instruments when observing from different vantage points. 

The spatial correlation and dependence of the linear fitting results are shown in Fig.~\ref{fig:spatial}. 
The correlation map shows very high values through the whole FoV with an average of 93\%, with the exception of the sunspot umbra, where the measured LoS velocities are generally lower than in the penumbra or quiet sun. 
This can be expected by the different algorithms used to retrieve the LoS velocity. 
\hmi\ uses the so-called MDI-like method for its LoS observable, whereas in \hrt\ the LoS velocity field is a result of the RTE inversion.  
Different algorithms can also infer the physical quantities at slightly different heights depending on their response functions. 
Such a possibility was not considered in \cite{2023-jesper}, because the MDI-like algorithm was used for both instruments. 
The use of different algorithms in the \hrt\ and \hmi\ pipelines for the retrieval of the velocities is possibly a further reason for the disagreement between the signals measured by these two instruments. 

The slope and offset obtained through the linear fit clearly show a dependence over the FoV. 
The offset variation is particularly strong in the left to right direction in the \hmi\ detector frame system. 
The only physical reason for such a gradient would come from the LSC, that already had been removed, but perhaps not entirely successfully. 
More advanced algorithms, such as that presented in \cite{2016-schuck}, can be used to improve the LSC correction, but in this work we applied the standard procedure based on well-established models. 
Another possibility is an instrumental origin of such a variation, which we investigated, but could not identify any clear cause due to instrumental properties of \hrt. 
Several tests using different calibration data have been tried and no significant difference is observed, such as improved flat fields or different prefilter models. 
In both instruments, several optical components along the respective optical paths show a FoV variation that needs to be fully and carefully characterized to remove any possible gradients in the two datasets.  
Also, there is no clear correlation with any of the LSC removed from the \hrt\ data that could cause such a signal, even if the models used for the LSC can yet leave residuals in the data \citep{2016-schuck}. 
Additionally, the amplitude of the variation is much higher than those of the LSC. 
It is also worth noting that the convective blueshift signal is highly sensitive to the spatial resolution of the instrument itself, which means that we expect a difference between \hrt\ and \hmi, but this is not included in this work (see Appendix~\ref{app:lsc}). 
The spatial variation can be treated more easily if the considered FoV is small and the removal of the offset and trends can be addressed with various methods \citep[e.g.][]{1997ApJ...474..810M,2011ARep...55..163G,2013ApJ...765...98W}, depending on the goal of the study. 
Such a scenario is clear also in the comparison of the penumbra in Fig.~\ref{fig:sunspot-scatter}. 
The spread of the distribution is substantially narrower also because of the smaller spread across the FoV and the value of the offset is also considerably closer to 0. 
On the contrary, the slope is similar to the one for the full FoV, indicating the robustness of the comparison and the weaker dependence over the FoV with respect to other quantities. 

As a last result, the Evershed flow within the sunspot penumbra is compared in Fig.~\ref{fig:sunspot}. 
The velocity signal displayed on an elliptical polar grid is used to isolate the red- and blue-shifted parts of the penumbra. 
The comparison between \hmi\ and \hrt, displayed in the right panel of Fig.~\ref{fig:sunspot}, shows very good agreement between the two instruments within the penumbra boundary and a small divergence in the quiet Sun comparable to the offset obtained with the linear fit shown in Fig.~\ref{fig:sunspot-scatter}. 
The difference measured right outside the penumbra does not change significantly along the rotation angle axis, showing a small dependence over such a small FoV. 
Please note that the velocities plotted in blue have to be flipped in sign before this invariance becomes clear. 
This is in agreement with the result displayed in the bottom panel of Fig.~\ref{fig:spatial}, meaning that this difference is due to the general offset between the two instruments. 
Stated differently, the penumbra itself shows higher velocity and smaller difference with respect to the quiet Sun, which could be due to how MILOS and the MDI-like algorithm infer the velocity in different physical conditions. 
The smaller difference in the penumbra than in the quiet Sun can also explain the smaller spread in Fig.~\ref{fig:sunspot-scatter} compared to Fig.~\ref{fig:scatter-full}. 

\section{Conclusion}\label{sec:end}
In summary, the agreement between the LoS velocity data products of \hrt\ and \hmi\ has been shown using a variety of analysis methods, providing quantitative supports for applications that combine the data from both instrument, from tracking of active regions to stereoscopic studies.
That being said, both instruments have calibration residuals that have to be treated with caution. 
There is not one univocal method to remove the offsets between the instruments without an actual absolute calibration. 
On the other hand, the difference in the rms can be accounted considering the dependence on the $\mu$ angle \citep[as shown in][]{2020ApJ...890..141O}. 
Such a procedure can be anyway only applied once the spatial resolution of the two instruments is comparable, as in this work, otherwise a higher spatial resolution will end in a higher velocity contrast \citep{2023LRSP...20....1J}. 
With such precautions, the data can also be combined pixel-by-pixel to measure 2-D flows on the solar surface from different vantage points. 
The excellent agreement between the instruments, as shown by the high correlation and by the slope consistently close to 1 as obtained by the linear fit on the scatter plot, will then allow the exploitation of the velocity datasets from \hrt\ and \hmi\ in multi-view configuration. 

\begin{acknowledgements}
We thank the anonymous referee for their comments and suggestions that improved the quality of this manuscript. Solar Orbiter is a space mission of international collaboration between ESA and NASA, operated by ESA. We are grateful to the ESA SOC and MOC teams for their support. The German contribution to SO/PHI is funded by the BMWi through DLR and by MPG central funds. The Spanish contribution is funded by AEI/MCIN/10.13039/501100011033/ and European Union ``NextGenerationEU/PRTR'' (RTI2018-096886-C5,  PID2021-125325OB-C5,  PCI2022-135009-2, PCI2022-135029-2) and ERDF ``A way of making Europe''; ``Center of Excellence Severo Ochoa'' awards to IAA-CSIC (SEV-2017-0709, CEX2021-001131-S). The French contribution is funded by CNES.
SDO is the first mission to be launched for NASA's Living With a Star (LWS) Program. The HMI data are courtesy of NASA/SDO and the HMI science team. This project has received funding from the European Research Council (ERC) under the European Union's Horizon 2020 research and innovation programme (grant agreement No. 101097844 — project WINSUN).
This research used version 6.0.6 (\href{https://doi.org/10.5281/zenodo.15690707}{10.5281/zenodo.15690707}) of the SunPy open source software package \citep{sunpy_community2020}.
\end{acknowledgements}

\bibliographystyle{aa}
\bibliography{bibfile}

\begin{appendix}
\section{Large scale velocity components}\label{app:lsc}
\begin{figure*}[htp]
    \centering
    \includegraphics[width=0.99\textwidth]{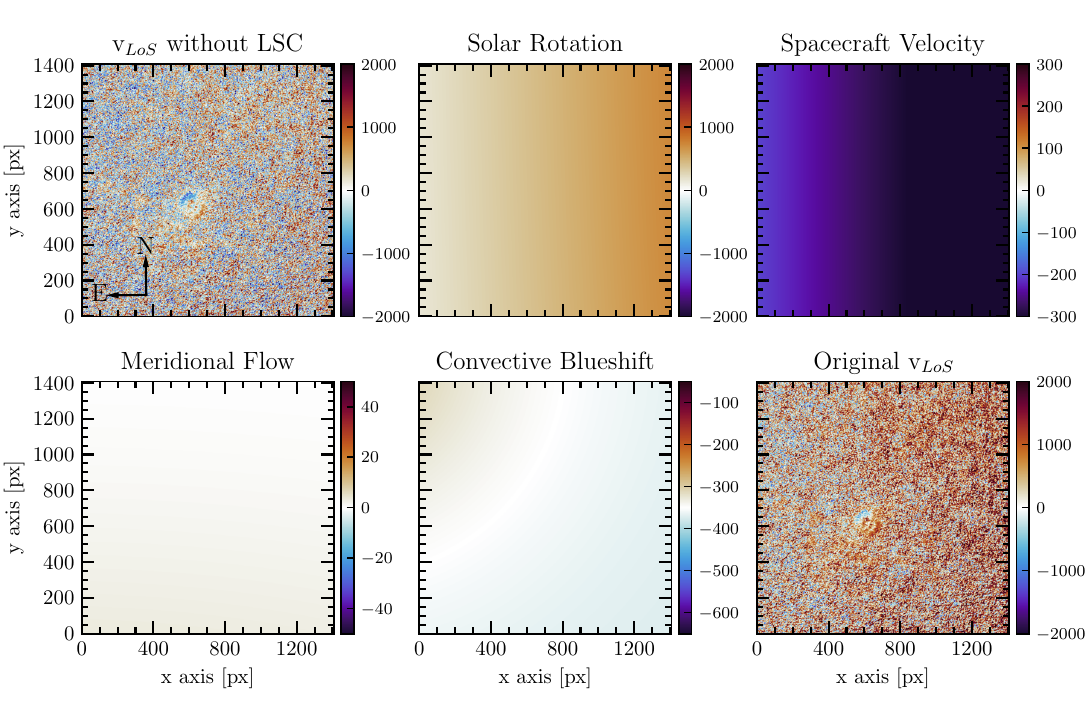}
    \caption{Top left panel: \hrt\ Dopplergram after the removal of the LSC. The black arrows indicate the north (N) and east (E) solar limb as in the \hrt\ detector frame. Top middle panel: differential solar rotation contribution. Top right panel: spacecraft velocity residuals. Bottom left panel: meridional flow contribution. Bottom middle panel: convective blueshift signal. Bottom right panel: original \hrt\ Dopplergram. The units are in m/s in every plot.}
    \label{fig:lsc}
\end{figure*}
The treatment of the large scale velocity components (LSC) in the solar photosphere and their contribution to the LoS velocity field has been extensively treated in many papers \citep[e.g.][]{2016-schuck,2021-castellanos}. 
Information about the solar coordinate system can be found in \cite{2006A&A...449..791T}. 
Here, for completeness, we summarize the equations used in our analysis. 
The solar differential rotation is treated according to \cite{2011-hathaway}:
\begin{equation}
    v_{rot}(\lambda) = (a + b\sin^2{\lambda} + c\sin^4{\lambda})\cos{\lambda}\,,
\end{equation}
where $\lambda$ is the solar latitude in heliographic coordinates and 
\begin{equation}
\begin{split}    
    a = &
    14.437\,^\circ/day
    \,,
    \\
    b = &
    -1.48\,^\circ/day
    \,,
    \\
    c = &
    -2.99\,^\circ/day
    \,.
\end{split}
\end{equation}
The projection along the LoS is given by \cite{2016-schuck}:
\begin{equation}
\begin{split}
    v_{rot}(\lambda)\cdot\hat{\eta}_{LoS} = \,&  v_{rot}(\lambda)\Big[ -\cos{B_0}\sin{\phi}\cos{\theta_\rho}+ \\
    & (\cos{\phi}\sin{\psi}-\sin{B_0}\sin{\phi}\cos{\psi})\sin{\theta_\rho} \Big]\, ,
\end{split}
\end{equation}
where $\hat{\eta}_{LoS}$ is the LoS vector in helioprojective coordinates, $B_0$ is the latitude of the center of the solar disk as seen by the observer (solar-B angle), $\phi$ is the solar longitude in heliographic coordinates (with 0\degree\ corresponding to the observer central meridian), $\theta_\rho$ is the helioprojective angle and $\psi$ is the position angle in heliocentric-radial coordinates. 

The LoS contribution of the residual spacecraft velocity component is:
\begin{equation}\label{eq:los}
    v_{S/C}(\theta_\rho,\psi) =  V_W\sin{\theta_\rho}\sin{\psi} - V_N \sin{\theta_\rho}\cos{\psi} +V_R\cos{\theta_\rho}\,,
\end{equation}
where ($V_R,V_W,V_N$) are the radial, westward, and northward components of the spacecraft velocity respectively. 
In the \hrt\ case, most of the contribution comes from the westward and northward velocities because the etalon can be tuned to compensate for the radial component of the spacecraft velocity.
Hence, we also subtract a constant value equal to 
\begin{equation}
    v_c(\lambda_c) = \frac{\lambda_{c}-\lambda_0}{\lambda_0}c\,,
\end{equation}
where $\lambda_c$ is the \hrt\ sampled wavelength corresponding to the core of the Fe~{\sc i} line, $\lambda_0$ is the rest wavelength, and $c$ is the speed of light. 

The meridional flow is also treated according to \cite{2011-hathaway}:
\begin{equation}
    v_M(\lambda) = (d\sin{\lambda}+e\sin^3{\lambda})\cos{\lambda}
\end{equation}
with 
\begin{equation}
    d=29.7\, m/s;\,e=-17.7\, m/s.
\end{equation}
The projection along the line of sight is equivalent to Equation~\ref{eq:los}. 

The variation from center to limb of the convective blueshift signal is adopted from \cite{2021-castellanos}. 
Despite the dependence of this effect on the spectral resolution \citep[see][]{stief-2019,boettcher-2019}, we are not able to fit the center to limb profile of this effect because \hrt\ is not a full-disk imager and we do not have continuous observations of the Sun \citep[e.g. as in][]{boettcher-2013}. 
This term is then modeled by \cite{2021-castellanos} as:
\begin{equation}
\begin{split}
    v_{BS}(\mu) = & 134-1179\mu-2029\mu^2+\\
    & +9112\mu^3-10409\mu^4+4096\mu^5\,m/s,
\end{split}
\end{equation}
where $\mu=\cos{\theta_\rho}$ is the cosine of the helioprojective angle. 

The last term is the gravitational redshift. 
This is a relativistic effect depending on the mass ($M_\odot$) and radius ($R_\odot$) of the Sun and on the distance ($D$) between the observer and the Sun:
\begin{equation}
    v_{GRS}(D) = \frac{GM_\odot}{c}\left( \frac{1}{R_\odot} - \frac{1}{D}\right),
\end{equation}
where $G$ is the gravitational constant. 
The values obtained for the two instruments are 628.8~m/s and 633.3~m/s for \hrt\ and \hmi\ respectively. 

Fig.~\ref{fig:lsc} shows all the components for \hrt\ except for the gravitational redshift which is constant. 

\end{appendix}
\end{document}